\documentclass[a4paper]{article}

\usepackage{INTERSPEECH2018}
\usepackage{graphicx}
\graphicspath{{./pdf/}}
\DeclareGraphicsExtensions{.pdf}
\usepackage[caption=false,font=footnotesize]{subfig}
\usepackage{amssymb,amsmath,bm}
\usepackage{mathrsfs}
\usepackage{enumitem}

\title{Speaker Embedding Extraction with Phonetic Information}
\name{Yi Liu$^1$, Liang He$^1$, Jia Liu$^1$, Michael T. Johnson$^2$}
\address{
  $^1$Tsinghua National Laboratory for Information Science and Technology, \\
  Department of Electronic Engineering, Tsinghua University, Beijing 100084, China \\
  $^2$Department of Electrical and Computer Engineering, University of Kentucky \\
  453 F. Paul Anderson Tower, Lexington, KY 40506-0046, USA}
\email{liu-yi15@mails.tsinghua.edu.cn, \{heliang, liuj\}@tsinghua.edu.cn, mike.johnson@uky.edu}

\begin{document}
\maketitle

\begin{abstract}
Speaker embeddings achieve promising results on many speaker verification tasks. Phonetic information, as an \mbox{important} component of speech, is rarely considered in the extraction of speaker embeddings. In this paper, we introduce phonetic information to the speaker embedding extraction based on the \mbox{x-vector} architecture. Two methods using phonetic vectors and multi-task learning are proposed. On the Fisher dataset, our best system outperforms the original x-vector approach by 20\% in EER, and by 15\%, 15\% in minDCF08 and minDCF10, respectively. Experiments conducted on NIST SRE10 further demonstrate the effectiveness of the proposed methods.
\end{abstract}

\section{Introduction}

During the last decade, the i-vector framework has become the dominant approach for speaker verification. By virtue of factor analysis and backend classifiers, the i-vector framework is able to model speaker characteristics and compensates for channel variability \cite{Dehak_ivector}. The main idea behind the framework is to represent a variable-length utterance with a fixed-length low-dimensional vector. These vectors can also be used in other areas such as speaker diarization \cite{Shum_diar} and speech synthesis \cite{Wan_tts}.

Recently, more attention has been drawn to the use of neural networks to extract speaker-discriminant vectors, which are known as \emph{speaker embeddings}. Early success of speaker embedding includes the d-vector, which was initially developed for text-dependent speaker verification \cite{Variani_dvector} and has been found to perform well in text-independent tasks \cite{Li_dvector}. To better capture speaker characteristics, recurrent neural networks (RNNs) \cite{Heigold_e2e}, convolutional neural networks (CNNs) \cite{Zhang_e2e} and many other neural network architectures \cite{Snyder_nin, Zhang_triplet, Li_deep} are used. Different loss criteria such as cross entropy \cite{Snyder_xvector}, triplet loss \cite{Zhang_triplet, Li_deep} and end-to-end loss \cite{Heigold_e2e, Zhang_e2e, Wan_e2e} have been investigated in recent publications. Speaker embeddings have outperformed conventional i-vectors in many conditions and are a promising new approach.

Current speaker embedding extraction only utilizes speaker labels and does not consider other information. However, speech signals are complex and are influenced by various factors. Two predominant components, phonetic content and speaker traits, are intermingled in the speech, representing \emph{what's said} and \emph{who's the speaker}. Other components include background noise and channel effects. This presents challenges to both speech and speaker recognition. In automatic speech recognition (ASR), speaker adaptation techniques are applied to eliminate the impact of diverse speakers and are effective at increasing accuracy \cite{Gales_mllr, Saon_adapt, Senior_adapt}. Similarly, phonetic independence is desirable in speaker recognition. Statistical models like joint factor analysis (JFA) \cite{Kenny_JFA} and i-vector can explicitly model the phonetic content \cite{Stafylakis_digit, Stafylakis_td}, while in neural models, phonetic vectors can be used as indicators to help deep neural networks (DNNs) find the more speaker-dependent information. In \cite{Li_factor}, the authors used output posteriors of an ASR network as the phonetic vectors and presented a form of speech factorization.

On the other hand, although the phonetic and speaker factors are different, they share some common information. For instance, the spectrum energy distribution and the trajectory of pitch are informative for speech and speaker recognition. Based on this fact, multi-task learning has been proposed in both areas \cite{Pironkov_ml, Chen_multitask, Tang_joint}. In multi-task learning, the speaker and phonetic discriminant networks share some hidden layers, and the networks predict speaker and phonetic labels at the same time.

However, there are some shortcomings in the previous phonetic vectors and multi-task learning based approaches:
\begin{enumerate}[leftmargin=*]
\item The phonetic vector is extracted from an independently trained ASR model, followed by the training of a speaker-discriminant network. This two-step procedure may lead to a sub-optimal result and the phonetic vector may be less efficient for speaker verification.
\item The existing methods for multi-task learning only perform in a frame-wise style, which means the speaker and phonetic classification are applied at the frame level. But speaker characteristics are often noisy at short-time frames, making it difficult to train the model.
\end{enumerate}

Therefore, in this paper, we propose two methods to combine the speaker embedding extraction with phonetic information. Joint training is introduced to improve the quality of phonetic vectors. In addition, a multi-task learning method operating at both frame and segment levels is presented. Experiments on Fisher and NIST SRE10 show that the proposed methods achieve better performance.

The organization of this paper is as follows. The speaker embedding we use in this paper is briefly introduced in Section 2. Section 3 describes the proposed methods to combine the phonetic information with the speaker embedding extraction. Our experimental setup and results are given in Section 4 and 5. The last section concludes the paper.

\section{Neural network speaker embedding}

\begin{figure}[t]
\centering
\includegraphics[width=0.8\linewidth]{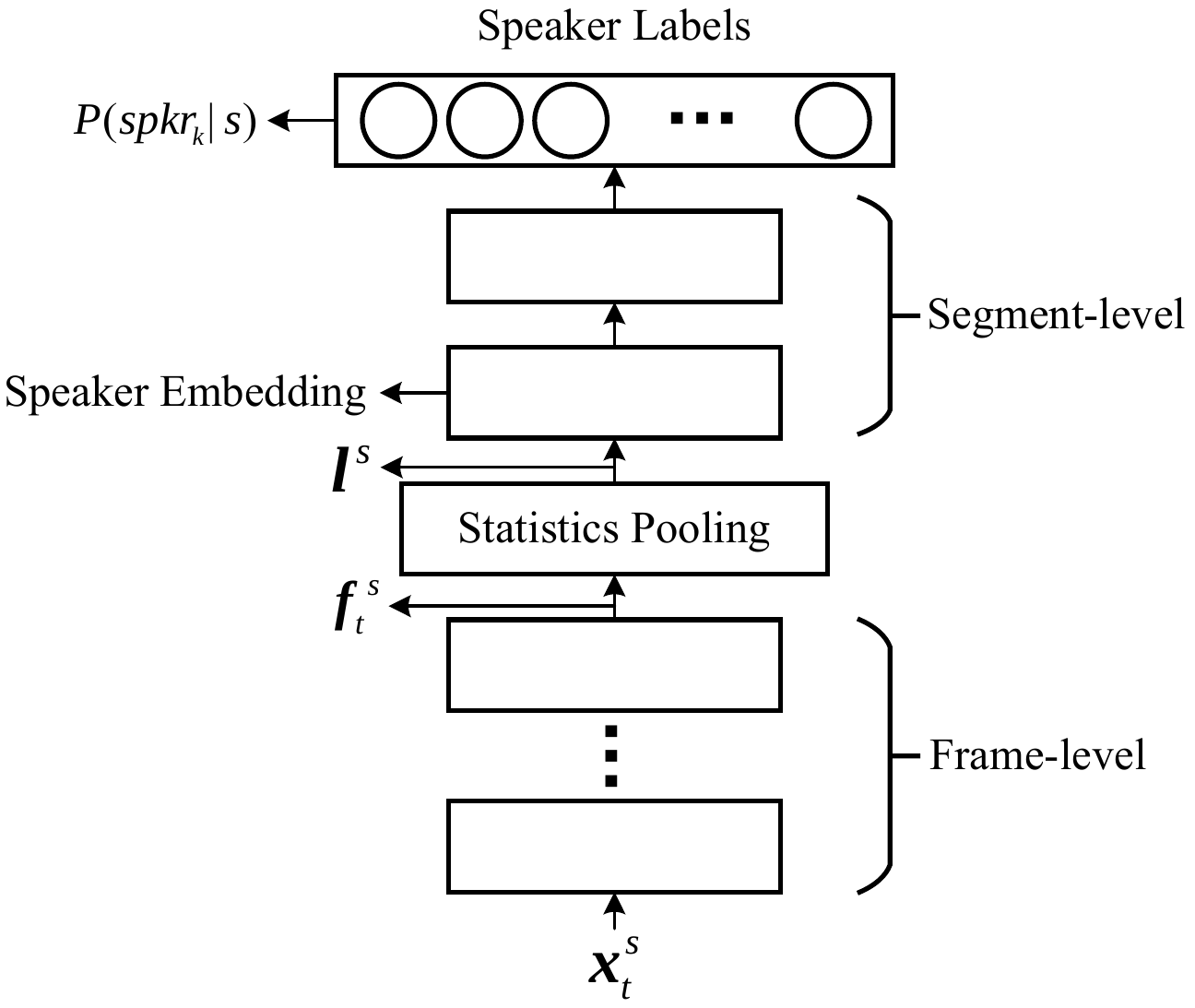}
\caption{The x-vector architecture for the speaker embedding extraction.}
\label{fig:x_vector}
\end{figure}
In this paper, we extract the speaker embedding based on the x-vector architecture \cite{Snyder_xvector}. As shown in Fig. \ref{fig:x_vector}, the network can be partitioned into frame and segment levels. The input $\vec{x}_t^s$ is the feature of frame $t$ in utterance $s$. After several hidden layers, a statistics pooling component is applied to all the frame-level activations $\vec{f}_t^s$, computes the mean and diagonal standard deviation, and reduces them to a segment vector $\vec{l}^s$. Fully-connected layers followed with a softmax layer are then used to predict the posterior of speaker $k$ with respect to utterance $s$. Given the parameters of the frame- and segment-level networks, $\vec{\theta}_f$ and $\vec{\theta}_l$, the posterior $P(\text{spkr}_k|s)$  is expressed as:
\begin{equation}\label{xvector_frame}
  \vec{f}_t^s = \mathcal{F}(\vec{x}_t^s, \vec{\theta}_f)
\end{equation}
\begin{equation}\label{stat_pooling}
  \vec{l}^s = \mathcal{P}(\vec{f}_1^s, \ldots, \vec{f}_{T_s}^s)
\end{equation}
\begin{equation}\label{xvector_utt}
  P(\text{spkr}_k|s) = \mathcal{F}(\vec{l}^s, \vec{\theta}_l)
\end{equation}
where $T_s$ is the utterance length and $\mathcal{F}(\cdot)$, $\mathcal{P}(\cdot)$ represent the forward and pooling operation, respectively.

After training, the output of a hidden layer at the segment level is extracted as the speaker embedding, termed the x-vector. Linear discriminant analysis (LDA) and probabilistic linear discriminant analysis (PLDA) \cite{Garcia_plda} are further applied to generate the verification scores.

\section{Phonetic information in speaker embedding extraction}

In this approach, the speaker embedding is extracted from a DNN. The phonetic information can be introduced by another neural acoustic model in ASR. The ASR network uses the same features but predicts the corresponding phonetic labels of the inputs. In contrast to the x-vector network, the ASR network works frame-by-frame. In this section, we propose two methods to combine these networks to extract the speaker embedding.

\subsection{Adapting x-vector with phonetic vectors}

\begin{figure}[t]
\centering
\includegraphics[width=\linewidth]{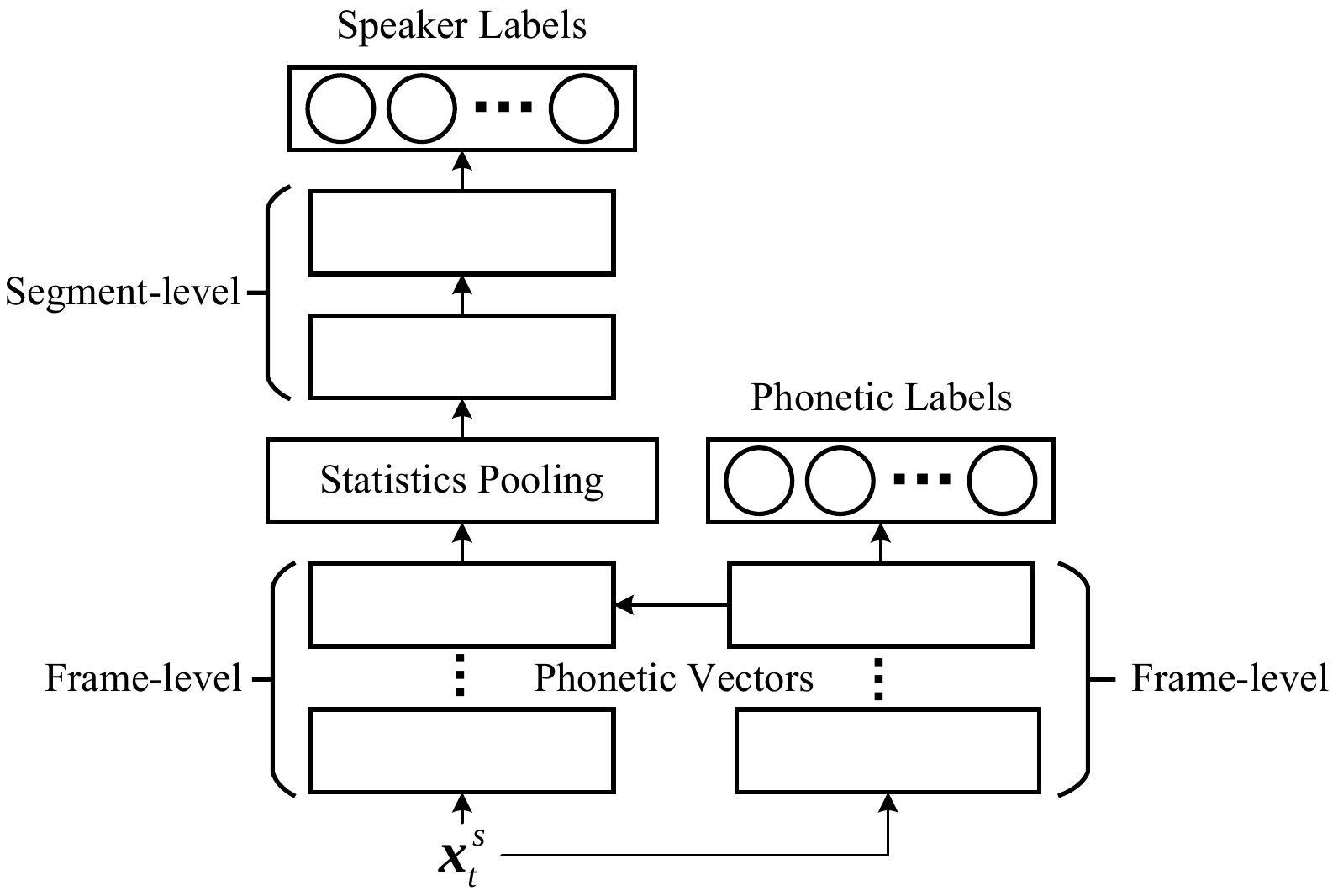}
\caption{The x-vector network with the auxiliary ASR network and phonetic vectors.}
\label{fig:phonetic_vector}
\end{figure}

Similar to speaker adaption based on speaker code \cite{Saon_adapt, Senior_adapt}, phonetic adaptation can also be applied using phonetic vectors. In our work, phonetic vectors are extracted from an ASR network and connected to the x-vector network as auxiliary inputs. For the initialization of the model, the ASR network with parameters $\{\vec{\theta}_a\}$ is first trained. The outputs of a bottleneck layer act as the phonetic vectors and the ASR network is attached to the x-vector network as Fig. \ref{fig:phonetic_vector}. Now, the frame-level activation $\vec{f}_t^s$ becomes:
\begin{equation}\label{xvector_frame}
  \vec{f}_t^s = \mathcal{F}(\vec{x}_t^s, \vec{\theta}_f, \vec{\theta}_a)
\end{equation}
During the x-vector training, the gradients back-propagate through the integrated network and update $\{\vec{\theta}_a, \vec{\theta}_f, \vec{\theta}_l\}$.

\subsection{Multi-task learning with shared layers}

\begin{figure}[t]
\centering
\includegraphics[width=\linewidth]{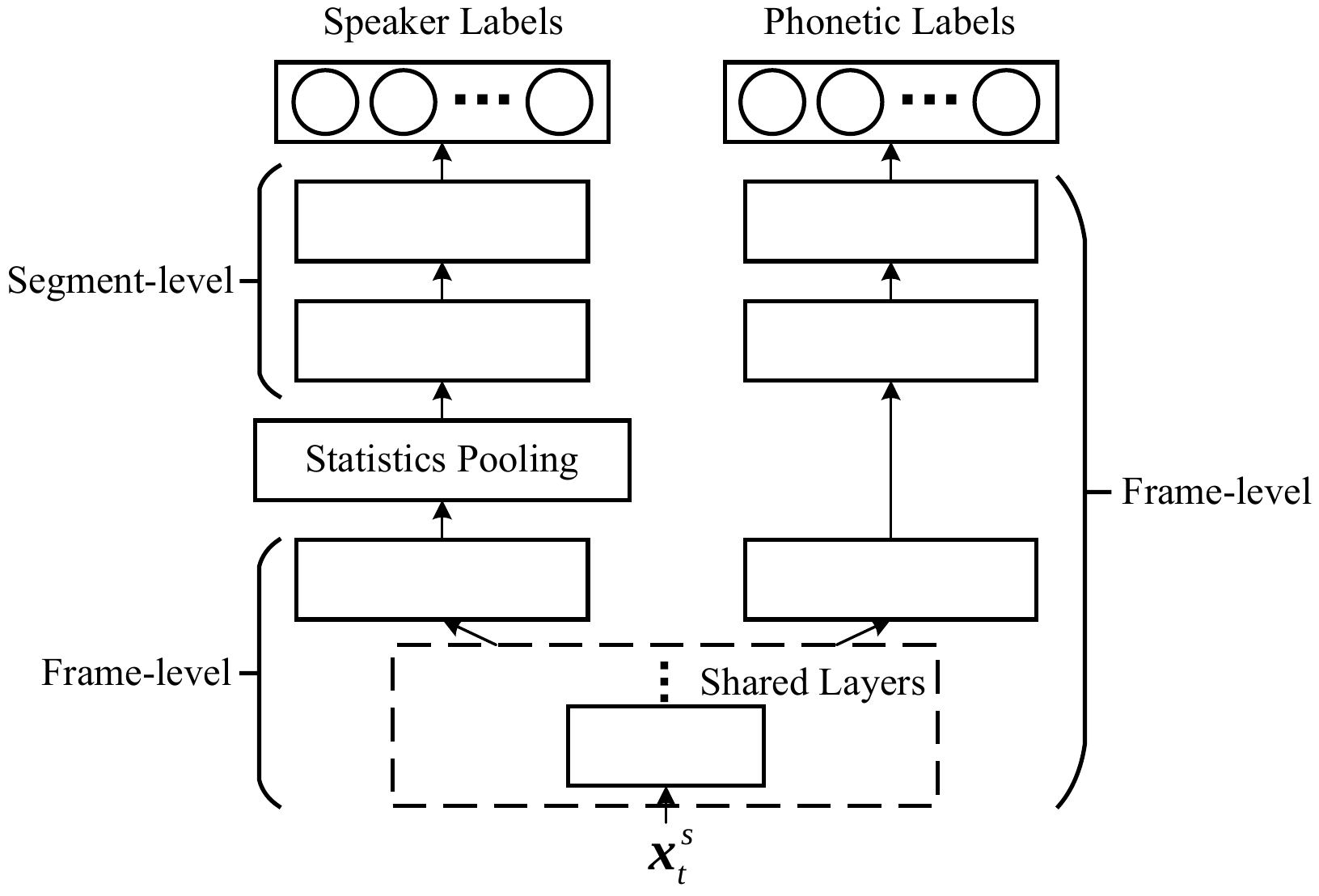}
\caption{Multi-task learning in a hybrid fashion at frame and segment levels.}
\label{fig:multitask_learning}
\end{figure}

Information which is vital for both speech and speaker recognition can be discovered by multi-task learning. Given a joint network, multi-task learning tries to perform well on both tasks. From a front-end processing perspective, the shared layers are a feature extractor which learns more informative features and suppresses nuisance noise in speech. From a view of model training, this improves the model generalization by introducing a regularization.

Since the speaker characteristics are noisy at the frame time-scale and tend to reside in longer segments, we propose a hybrid multi-task learning.  In contrast to existing works, in our method only the frame-level part of the x-vector network is shared with the ASR network. The phonetic classification is done at the frame level, while the speaker labels are classified at the segment level. The architecture is shown in Fig. \ref{fig:multitask_learning}.

In this approach, the parameters are split into $\{\vec{\theta}_s, \vec{\theta}'_a, \vec{\theta}'_f, \vec{\theta}_l\}$ in the new model, where $\vec{\theta}_s$ relates to the shared layers, and $\vec{\theta}'_a$, $\vec{\theta}'_f$ denote the remaining part of the ASR and x-vector networks at the frame level. The activation $\vec{f}_t^s$ changes to:
\begin{equation}\label{xvector_frame}
  \vec{f}_t^s = \mathcal{F}(\vec{x}_t^s, \vec{\theta}_s, \vec{\theta}'_f)
\end{equation}

A strategy similar to the training of multilingual acoustic models is applied \cite{Sercu_ml}. There are two types of training examples in our model, i.e., the speaker and phonetic examples. Given input features, the speaker examples only contain the speaker labels while the phonetic examples contain the corresponding phonetic units. During the training process, these examples are merged into different mini-batches. When the speaker examples are used, we update $\{\vec{\theta}_s, \vec{\theta}'_f, \vec{\theta}_l\}$. When the phonetic examples are used, $\{\vec{\theta}_s, \vec{\theta}'_a\}$ are updated. It is possible to choose different learning rates for these tasks, but for simplicity we  keep them the same in the experiments.

This training strategy is flexible. By introducing other data designed for ASR, the multi-task learning can be used even if the original dataset does not have any phonetic transcriptions.

\section{Experimental setup}

\subsection{Dataset}

To investigate the performance of the speaker embedding extraction with phonetic information, we have run experiments on two datasets: Fisher \cite{Cieri_fisher} and NIST SRE10 \cite{Martin_sre10}. Fisher is speaker and ASR transcribed, which provides an ideal condition to assess our methods. On NIST SRE10, however, only speaker labels are available. To employ phonetic information in this case, we need to introduce an out-of-domain ASR dataset. The details of the two datasets are given as follows.
\begin{itemize}[leftmargin=*]
\item \emph{Fisher} is manually partitioned into training and evaluation subsets. We randomly select 95167 segments (sampled from 13390 utterances) as the training set. There are 5000 speakers in the 172h training set. The duration of each segment ranges from 3s to 15s. The evaluation set contains 1000 speakers which do not overlap with the training set. The enrollment for each speaker consists of 10 segments (about 30 seconds in total). The test data contains 3000 segments (3 segments per person and 3 seconds per segment), forming 3M target and non-target trials.
\item \emph{NIST SRE10} core-extended and 10s-10s condition 5 are used in our experiments. NIST SRE 2004-2008 telephone excerpts, Switchboard Phase II Part 1/2/3 and Cellular Part 1/2 are used as the training set. This represents 5524 hours of data and comprises 6374 speakers, 64742 utterances. To introduce phonetic information, the 318-hour out-of-domain Switchboard-I is used.
\end{itemize}
For both datasets, the validation set consists of 1000 segments randomly selected from the training set.

All models in our experiments are gender-independent, and the results are reported on the male and female pooled trials. Since we do not have enough space to illustrate the detection error tradeoff (DET) plots, equal error rate (EER), minimum detection cost function from NIST SRE08 (minDCF08) \cite{Martin_sre08} and SRE10 (minDCF10) \cite{Martin_sre10} are presented to investigate the performance across different operation points.

\subsection{Baseline i-vector system}

A standard i-vector system is used as a comparison to the x-vector and our proposed methods. The feature is 20-dimension static MFCCs with delta and delta-delta. Cepstral mean normalization (CMN) and energy-based voice active detection (VAD) are applied. We use the same features in all experiments.

The 2048-mixture universal background model (UBM) and the 600-dimension i-vector extractor are trained using the training data. LDA is applied to reduce the dimension of i-vector to 200 prior to PLDA scoring.

\subsection{X-vector architecture}

The frame-level part of the x-vector network is a 5-layer time-delay neural network (TDNN) \cite{Peddinti_tdnn}. The input of each layer is the sliced output of the previous layer. The slicing parameter is: $\{t-2,t-1,t,t+1,t+2\}$, $\{t-2,t,t+2\}$, $\{t-3,t,t+3\}$, $\{t\}$, $\{t\}$. It has 512 nodes in layer 1 to 4, and the 5-th layer has 1500 nodes. The segment-level part is a 2-layer fully-connected network with 512 nodes per layer. The output is predicted by softmax and the size is the same as the number of speakers. 150-dimensionl LDA and PLDA scoring are trained and applied. Refer to \cite{Snyder_xvector} for more details.

\subsection{X-vector extraction with phonetic information}

In our experiments, we use senones as the phonetic units to train the ASR networks. The number of senones is 2366 for Fisher, and 3854 for Switchboard-I. The senone transcriptions is generated by GMM-HMM forced alignment.

For adaptation based on phonetic vectors, a 5-layer TDNN ASR network is first trained until convergence. The last layer is a bottleneck layer with 128 nodes, and other layers consists of 512 nodes. The output of the bottleneck layer is connected to the 5-th layer of the x-vector network. The combined network is then jointly optimized. The learning rate scale factor to fine-tune the ASR network is 0.2.

In multi-task learning, the ASR network has the same architecture with the x-vector network but without the statistics pooling component, since they share common layers. The batch sizes for speaker and phonetic examples are 64 and 256, respectively.

Our systems are implemented with Kaldi toolkit \cite{Povey_kaldi} and will be open-sourced \footnote{yiliu.org.cn}.

\section{Results}

\subsection{Fisher}

\begin{table}[thbp]
\caption{Results on Fisher dataset. PV denotes the speaker embedding with phonetic vectors and MT-$n$ is the multi-task learning sharing $n$ layers. }
\label{table:result_fisher}
\centering
\begin{tabular}{|c|c|c|c|}
\hline
Systems & EER(\%) & minDCF08 & minDCF10 \\
\hline
\hline
i-vector & 2.10 & 0.0093 & 0.3347 \\
\hline
x-vector & 1.73 & 0.0086 & 0.3627 \\
\hline
\hline
PV (no fine-tuning) & 1.63 & 0.0089 & 0.3607 \\
\hline
PV (fine-tuning) & \textbf{1.60} & \textbf{0.0076} & \textbf{0.3413} \\
\hline
\hline
MT-1 & 1.73 & 0.0082 & 0.3487 \\
\hline
MT-2 & 1.53 & 0.0078 & 0.3170 \\
\hline
MT-3 & 1.57 & 0.0072 & 0.3080 \\
\hline
MT-4 & \textbf{1.39} & \textbf{0.0073} & \textbf{0.3087} \\
\hline
MT-5 & 1.50 & 0.0076 & 0.3323 \\
\hline
\end{tabular}
\end{table}

The results of i-vector and x-vector are first compared on Fisher dataset. From Table \ref{table:result_fisher}, it can be seen that x-vector outperforms i-vector in EER and minDCF08, while i-vector is better in minDCF10. The performance of speaker embeddings with phonetic vectors is shown in the second part of Table \ref{table:result_fisher}. Using phonetic vectors extracted from the ASR network without fine-tuning, the speaker embedding performs slightly better than x-vector. With the network fine-tuning, the performance is further improved.

Next, the proposed hybrid multi-task learning methods are evaluated. The results of sharing different layers are illustrated in the third part of Table \ref{table:result_fisher}. When the number of shared layers increases, the performance first improves and then slightly degrades. Sharing 4 frame-level layers achieves the best result in our experiments. This gives a 20\%, 15\% and 15\% improvement over the original x-vector in EER, minDCF08 and minDCF10, respectively.

\begin{figure}[t]
\centering
\includegraphics[width=0.8\linewidth]{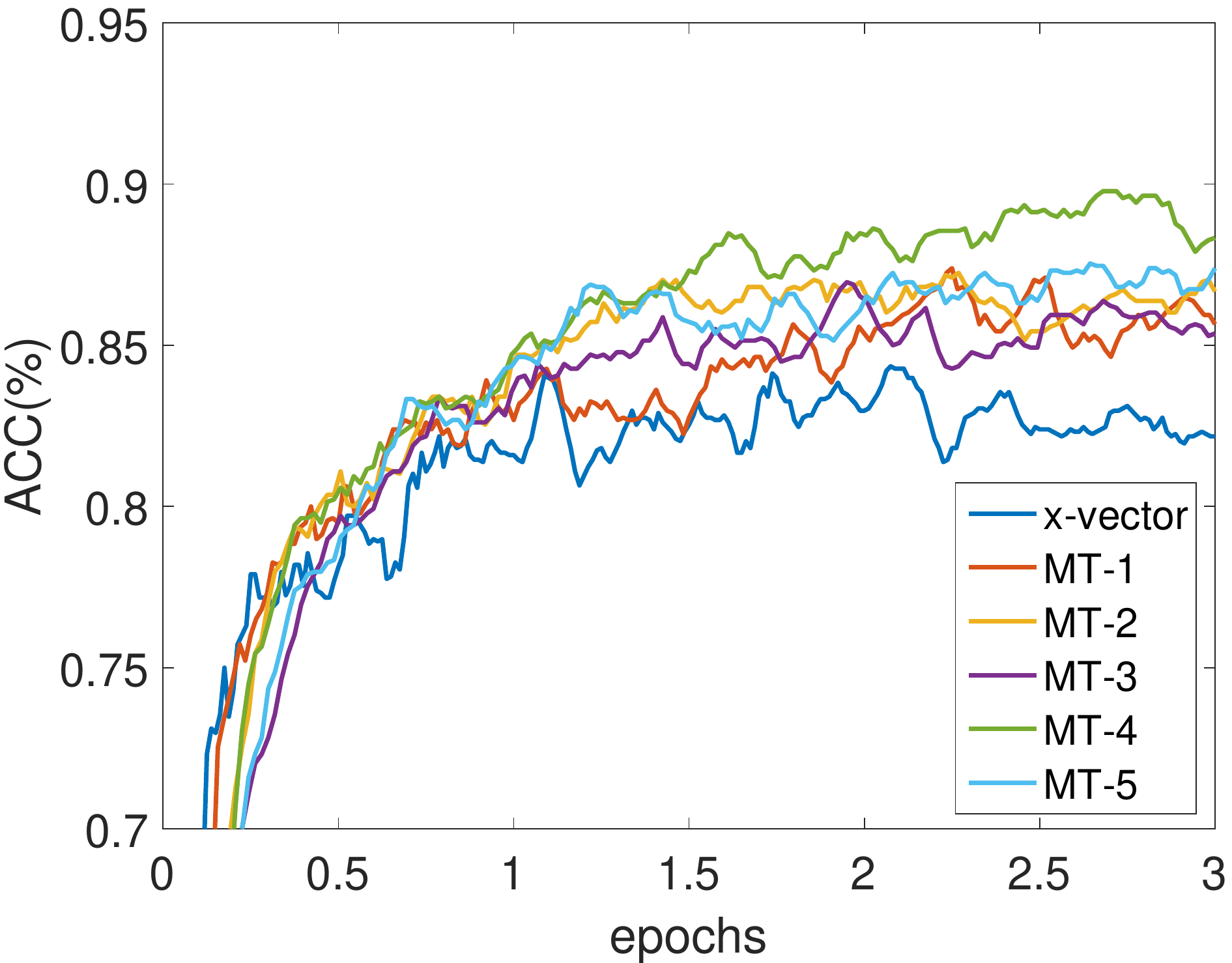}
\caption{Speaker classification accuracy on Fisher validation set vs epochs trained.}
\label{fig:acc_dev}
\end{figure}

As stated before, the ASR network in the multi-task learning can be seen as a regularization. Fig. \ref{fig:acc_dev} shows the speaker classification accuracy on the validation set using different settings. From this figure, we find that sharing 4 layers achieves the highest validation accuracy, while the unmodified x-vector performs the worst. The models trained by multi-task learning generalize better to unseen data. This is consistent with the results of speaker verification in Table \ref{table:result_fisher}.

\subsection{NIST SRE10}

The speaker and ASR data are perfectly matched on the Fisher set. This is not the case for NIST SRE10. We have to involve the out-of-domain Switchboard-I to provide the phonetic information for the speaker embedding extraction.

\begin{table}[thbp]
\caption{Results on NIST SRE10 core-extended condition 5. PV denotes the speaker embedding with phonetic vectors and MT-$n$ is the multi-task learning sharing $n$ layers. }
\label{table:result_sre_coreext}
\centering
\begin{tabular}{|c|c|c|c|}
\hline
Systems & EER(\%) & minDCF08 & minDCF10 \\
\hline
\hline
i-vector & 2.19 & 0.0120 & 0.4373 \\
\hline
x-vector & 2.23 & 0.0124 & 0.4593 \\
\hline
\hline
PV (no fine-tuning) & 1.84 & 0.0101 & 0.4245\\
\hline
PV (fine-tuning) & \textbf{1.61} & \textbf{0.0093} & \textbf{0.3643} \\
\hline
\hline
MT-1 & 1.78 & 0.0103 & 0.3861 \\
\hline
MT-2 & 1.76 & 0.0101 & 0.3717 \\
\hline
MT-3 & \textbf{1.59} & \textbf{0.0102} & \textbf{0.3700} \\
\hline
MT-4 & 1.90 & 0.0101 & 0.3652 \\
\hline
\end{tabular}
\end{table}

\begin{table}[thbp]
\caption{Results on NIST SRE10 10s-10s condition 5. PV denotes the speaker embedding with phonetic vectors and MT-$n$ is the multi-task learning sharing $n$ layers. }
\label{table:result_sre_10s}
\centering
\begin{tabular}{|c|c|c|c|}
\hline
Systems & EER(\%) & minDCF08 & minDCF10 \\
\hline
\hline
i-vector & 10.46 & 0.0533 & 0.9817 \\
\hline
x-vector & 9.15 & 0.0479 & 0.9093 \\
\hline
\hline
PV (no fine-tuning) & 9.01 & 0.0453 & 0.9093 \\
\hline
PV (fine-tuning) & \textbf{8.45} & \textbf{0.0424} & \textbf{0.8828} \\
\hline
\hline
MT-1 & 9.14 & 0.0446 & 0.8809 \\
\hline
MT-2 & \textbf{8.05} & \textbf{0.0430} & 0.9130 \\
\hline
MT-3 & 8.81 & 0.0468 & \textbf{0.8342} \\
\hline
MT-4 & 10.07 & 0.0505 & 0.9103 \\
\hline
\end{tabular}
\end{table}

Table \ref{table:result_sre_coreext} presents the results of different methods on the NIST SRE10 core-extended condition 5. When the utterance duration is long enough, i-vector achieves a good performance and outperforms x-vector across all the three operation points. Using phonetic vectors, the speaker embedding again performs better than i-vector. We note that the phonetic vector with fine-tuning outperforms the one without fine-tuning by 13\%, 8\% and 14\% in EER, minDCF08 and minDCF10. The multi-task learning approach is also effective. With 3 layers shared, multi-task learning improves the performance of x-vector by 29\% in EER, 18\% in minDCF08 and 19\% in minDCF10.

As shown in Table \ref{table:result_sre_10s}, the speaker embeddings with phonetic vectors and multi-task learning also achieve better results than i-vector and x-vector on the NIST SRE10 10s-10s condition 5. The improvements are relatively smaller than the previous conditions. Using phonetic vectors with fine-tuning performs the best in minDCF08 while the multi-task learning sharing different layers is optimal for EER and minDCF10.

On the NIST SRE10 dataset, the optimal number of shared layers in multi-task learning is generally less than that for Fisher. The reason is that NIST SRE10 has more than 5000 hours of data to train the x-vector network. It is thus easier to learn the speaker characteristics from the source and is less likely to overfit. Moreover, the phonetic information comes from the out-of-domain $\sim$300h Switchboard-I dataset, limiting its contribution to the training. Even under this condition, the speaker embedding still benefits from the phonetic information.

\section{Conclusions}

In this paper, we add phonetic information to the speaker embedding extraction based on phonetic vectors and multi-task learning. The phonetic vector is jointly optimized with the x-vector network and provides auxiliary information to adapt the speaker embedding. A novel multi-task learning, working at both frame and segment levels, is then proposed. The shared layers extract informative features, making the network more robust and less likely to overfit. The speaker embeddings extracted with phonetic vectors and multi-task learning substantially outperform the baseline i-vector and x-vector on the Fisher dataset. On NIST SRE10, using the mismatched Switchboard-I data, speaker embedding still shows benefits from adding the phonetic information. By carefully designing the network architecture, the proposed methods can be applied to other end-to-end speaker verification systems.

In the future, we will improve the method when only out-of-domain ASR data is available, and develop an architecture to combine the phonetic vector approach with multi-task learning.

\section{Acknowledgements}
The work is supported by National Natural Science Foundation of China under Grant No. 61370034, No. 61403224 and No. 61273268.

\bibliographystyle{IEEEtran}
\bibliography{mybib}

\begin{thebibliography}{10}
\providecommand{\url}[1]{#1}
\csname url@samestyle\endcsname
\providecommand{\newblock}{\relax}
\providecommand{\bibinfo}[2]{#2}
\providecommand{\BIBentrySTDinterwordspacing}{\spaceskip=0pt\relax}
\providecommand{\BIBentryALTinterwordstretchfactor}{4}
\providecommand{\BIBentryALTinterwordspacing}{\spaceskip=\fontdimen2\font plus
\BIBentryALTinterwordstretchfactor\fontdimen3\font minus
  \fontdimen4\font\relax}
\providecommand{\BIBforeignlanguage}[2]{{%
\expandafter\ifx\csname l@#1\endcsname\relax
\typeout{** WARNING: IEEEtran.bst: No hyphenation pattern has been}%
\typeout{** loaded for the language `#1'. Using the pattern for}%
\typeout{** the default language instead.}%
\else
\language=\csname l@#1\endcsname
\fi
#2}}
\providecommand{\BIBdecl}{\relax}
\BIBdecl

\bibitem{Dehak_ivector}
N.~Dehak, P.~J. Kenny, R.~Dehak, P.~Dumouchel, and P.~Ouellet, ``Front-end
  factor analysis for speaker verification,'' \emph{IEEE Transactions on Audio,
  Speech, and Language Processing}, vol.~19, no.~4, pp. 788--798, 2011.

\bibitem{Shum_diar}
S.~Shum, N.~Dehak, E.~Chuangsuwanich, D.~Reynolds, and J.~Glass, ``Exploiting
  intra-conversation variability for speaker diarization,'' in \emph{Proc.
  {INTERSPEECH}}, 2011, pp. 945--948.

\bibitem{Wan_tts}
M.~Wan, G.~Degottex, and M.~J. Gales, ``Integrated speaker-adaptive speech
  synthesis,'' in \emph{Proc. IEEE Automatic Speech Recognition and
  Understanding Workshop (ASRU)}, 2017, pp. 705--711.

\bibitem{Variani_dvector}
E.~Variani, X.~Lei, E.~McDermott, I.~L. Moreno, and J.~Gonzalez-Dominguez,
  ``Deep neural networks for small footprint text-dependent speaker
  verification,'' in \emph{Proc. IEEE {ICASSP}}, May 2014, pp. 4052--4056.

\bibitem{Li_dvector}
L.~Li, Y.~Chen, Y.~Shi, Z.~Tang, and D.~Wang, ``Deep speaker feature learning
  for text-independent speaker verification,'' in \emph{Proc. {INTERSPEECH}},
  2017, pp. 1542--1546.

\bibitem{Heigold_e2e}
G.~Heigold, I.~Moreno, S.~Bengio, and N.~Shazeer, ``End-to-end text-dependent
  speaker verification,'' in \emph{Proc. IEEE {ICASSP}}, March 2016, pp.
  5115--5119.

\bibitem{Zhang_e2e}
S.-X. Zhang, Z.~Chen, Y.~Zhao, J.~Li, and Y.~Gong, ``End-to-end attention based
  text-dependent speaker verification,'' in \emph{Proc. IEEE Spoken Language
  Technology Workshop (SLT)}, 2016, pp. 171--178.

\bibitem{Snyder_nin}
D.~Snyder, P.~Ghahremani, D.~Povey, D.~Garcia-Romero, Y.~Carmiel, and
  S.~Khudanpur, ``Deep neural network-based speaker embeddings for end-to-end
  speaker verification,'' in \emph{Proc. IEEE Spoken Language Technology
  Workshop (SLT)}, 2016, pp. 165--170.

\bibitem{Zhang_triplet}
C.~Zhang and K.~Koishida, ``End-to-end text-independent speaker verification
  with triplet loss on short utterances,'' in \emph{Proc. {INTERSPEECH}}, 2017,
  pp. 1487--1491.

\bibitem{Li_deep}
C.~Li, X.~Ma, B.~Jiang, X.~Li, X.~Zhang, X.~Liu, Y.~Cao, A.~Kannan, and Z.~Zhu,
  ``Deep speaker: An end-to-end neural speaker embedding system,'' \emph{arXiv
  preprint arXiv:1705.02304}, 2017.

\bibitem{Snyder_xvector}
D.~Snyder, D.~Garcia-Romero, D.~Povey, and S.~Khudanpur, ``Deep neural network
  embeddings for text-independent speaker verification,'' in \emph{Proc.
  {INTERSPEECH}}, 2017, pp. 999--1003.

\bibitem{Wan_e2e}
L.~Wan, Q.~Wang, A.~Papir, and I.~L. Moreno, ``Generalized end-to-end loss for
  speaker verification,'' \emph{arXiv preprint arXiv:1710.10467}, 2017.

\bibitem{Gales_mllr}
M.~J.~F. Gales, ``Maximum likelihood linear transformations for hmm-based
  speech recognition,'' \emph{Computer Speech \& Language}, vol.~12, no.~2, pp.
  75--98, 1998.

\bibitem{Saon_adapt}
G.~Saon, H.~Soltau, D.~Nahamoo, and M.~Picheny, ``Speaker adaptation of neural
  network acoustic models using i-vectors,'' in \emph{Proc. IEEE Automatic
  Speech Recognition and Understanding Workshop (ASRU)}, 2013, pp. 55--59.

\bibitem{Senior_adapt}
A.~Senior and I.~Lopez-Moreno, ``Improving {DNN} speaker independence with
  i-vector inputs,'' in \emph{Proc. IEEE {ICASSP}}, 2014, pp. 225--229.

\bibitem{Kenny_JFA}
P.~Kenny, ``Joint factor analysis of speaker and session variability: Theory
  and algorithms,'' \emph{CRIM, Montreal,(Report) CRIM-06/08-13}, vol.~14, pp.
  28--29, 2005.

\bibitem{Stafylakis_digit}
T.~Stafylakis, P.~Kenny, J.~Alam, and M.~Kockmann, ``{JFA} for speaker
  recognition with random digit strings,'' in \emph{Proc. {INTERSPEECH}}, 2015,
  pp. 190--194.

\bibitem{Stafylakis_td}
T.~Stafylakis, P.~Kenny, M.~J. Alam, and M.~Kockmann, ``Speaker and channel
  factors in text-dependent speaker recognition,'' \emph{IEEE/ACM Transactions
  on Audio, Speech, and Language Processing}, vol.~24, no.~1, pp. 65--78, 2016.

\bibitem{Li_factor}
L.~Li, D.~Wang, Y.~Chen, Y.~Shi, Z.~Tang, and T.~F. Zheng, ``Deep factorization
  for speech signal,'' \emph{arXiv preprint arXiv:1803.00886}, 2018.

\bibitem{Pironkov_ml}
G.~Pironkov, S.~Dupont, and T.~Dutoit, ``Speaker-aware long short-term memory
  multi-task learning for speech recognition,'' in \emph{Proc. IEEE Signal
  Processing Conference (EUSIPCO)}, 2016, pp. 1911--1915.

\bibitem{Chen_multitask}
N.~Chen, Y.~Qian, and K.~Yu, ``Multi-task learning for text-dependent speaker
  verification,'' in \emph{Proc. {INTERSPEECH}}, 2015, pp. 185--189.

\bibitem{Tang_joint}
Z.~Tang, L.~Li, D.~Wang, and R.~Vipperla, ``Collaborative joint training with
  multitask recurrent model for speech and speaker recognition,''
  \emph{IEEE/ACM Transactions on Audio, Speech, and Language Processing},
  vol.~25, no.~3, pp. 493--504, 2017.

\bibitem{Garcia_plda}
D.~Garcia-Romero and C.~Y. Espy-Wilson, ``Analysis of i-vector length
  normalization in speaker recognition systems,'' in \emph{Proc.
  {INTERSPEECH}}, 2011, pp. 256--259.

\bibitem{Sercu_ml}
T.~Sercu, C.~Puhrsch, B.~Kingsbury, and Y.~LeCun, ``Very deep multilingual
  convolutional neural networks for lvcsr,'' in \emph{Proc. IEEE {ICASSP}},
  2016, pp. 4955--4959.

\bibitem{Cieri_fisher}
C.~Cieri, D.~Miller, and K.~Walker, ``The {Fisher} corpus: a resource for the
  next generations of speech-to-text,'' in \emph{LREC}, vol.~4, 2004, pp.
  69--71.

\bibitem{Martin_sre10}
A.~F. Martin and C.~S. Greenberg, ``The {NIST} 2010 speaker recognition
  evaluation,'' in \emph{Proc. {INTERSPEECH}}, 2010, pp. 2726--2729.

\bibitem{Martin_sre08}
------, ``{NIST} 2008 speaker recognition evaluation: Performance across
  telephone and room microphone channels,'' in \emph{Proc. {INTERSPEECH}},
  2009, pp. 2579--2582.

\bibitem{Peddinti_tdnn}
V.~Peddinti, D.~Povey, and S.~Khudanpur, ``A time delay neural network
  architecture for efficient modeling of long temporal contexts,'' in
  \emph{Proc. {INTERSPEECH}}, 2015.

\bibitem{Povey_kaldi}
D.~Povey, A.~Ghoshal, G.~Boulianne, L.~Burget, O.~Glembek, N.~Goel,
  M.~Hannemann, P.~Motlicek, Y.~Qian, P.~Schwarz, and Others, ``The kaldi
  speech recognition toolkit,'' in \emph{Proc. IEEE Automatic Speech
  Recognition and Understanding Workshop (ASRU)}, 2011.

\end{thebibliography}

\end{document}